# Dielectric-tensor reconstruction of highly scattering birefringent samples

Hervé Hugonnet[1,2,†], Moosung Lee[1,2,†], Seungwoo Shin[1,3], and YongKeun Park[1,2,4*]

[1] Department of Physics, Korea Advanced Institute of Science and Technology (KAIST), Daejeon 34141, South Korea;

[2] KAIST Institute for Health Science and Technology, KAIST, Daejeon 34141, South Korea;

[3] Current affiliation: Department of Physics, University of California at Santa Barbara, Santa Barbara, California 93106, USA

[4] Tomocube Inc., Daejeon 34109, South Korea

[†]These authors equally contributed to the work.

*corresponding authors: Y.K.P (yk.park@kaist.ac.kr)

**ABSTRACT**

Many important microscopy samples, such as liquid crystals, biological tissue, or starches, are birefringent in nature. They scatter light differently depending on the light polarization and molecular orientations. The complete characterization of a birefringent sample is a challenging task because its $3 \times 3$ dielectric tensor must be reconstructed at every three-dimensional position. Moreover, obtaining a birefringent tomogram is more arduous for thick samples, where multiple light scattering should also be considered. In this study, we developed a new dielectric tensor tomography algorithm that enables full characterization of highly scattering birefringent samples by solving the vectoral inverse scattering problem considering multiple light scattering. We proposed a discrete image-processing theory to compute the error backpropagation of vectorially diffracting light. Finally, our theory was experimentally demonstrated using both synthetic and biologically birefringent samples.

# INTRODUCTION

Birefringence occurs during the scattering of polarization-dependent light by optically anisotropic materials. Samples with heterogeneous birefringence frequently emerge in soft matter physics and biophysics when molecules locally align their orientations because of complex many-body interactions[1-7]. Quantitative imaging of three-dimensional (3D) birefringence reveals the underlying structural order of these molecules, which may comprise molecules or crystals in biological tissues. However, these birefringent materials are mostly translucent and exhibit low contrast under conventional microscopic methods, which results in inaccurate characterization when using conventional bright-field microscopes[8-12].

Quantitative phase imaging (QPI) exploits the refractive index (RI) as an intrinsic quantitative imaging contrast and has been successfully deployed in many applications[13-15] where translucent samples should be imaged. Previously 2D polarization sensitive QPI techniques has been reported and utilized for measuring polarization sensitive phase retardance or Jones matrix information[8-11]. Optical diffraction tomography (ODT) reconstructs the 3D refractive index (RI) distribution of a sample[16-19]. Extended from the ODT principle, recent progress has enabled direct 3D imaging of sample birefringence[20-22] and even direct measurement of the full 3D dielectric tensor[23,24]. However, these methods were all based on the weak scattering approximation in order to obtain an analytical solution for the inverse scattering program. This weak scattering approximation limits the general applications to complex materials with strong birefringence.

This study solved the inverse scattering problem beyond the weak scattering regime to reconstruct the dielectric tensor distribution of highly scattering 3D birefringent samples [Fig. 1a]. We specifically extended our previous discrete model of error backpropagation from scalar wave theory to a vectoral one[25]. The advantages of the proposed theory were experimentally demonstrated via the reconstruction of the 3D dielectric tensor distribution of synthetic liquid crystal droplets and biological starch grains. The dielectric tensor can be decomposed using singular value decomposition into three orthogonal polarization directions and three refractive indices for each of the corresponding polarizations. Consequently, the director and amount of birefringence can also be quantified [Fig. 1b]. The polarization direction corresponding to the highest refractive index corresponds to the underlying molecular alignment and is referred to as the director. Contrary to previous methods[25-28], the proposed model does not require strict assumptions, such as weak scattering or scalar diffraction. Moreover, it inversely solves Maxwell's equations to fully characterize birefringent dielectric materials.

## Results

**Vectoral diffraction theory in discrete space**

Understanding the light scattering by a birefringent object is essential for the development of advanced dielectric tensor imaging techniques. Consider a sample with dielectric tensor distribution, $\ddot{\varepsilon}(\vec{r})$, which scatters the incident polarized light field, $\vec{E}_{in}(\vec{r})$. For a dielectric material illuminated with monochromatic light, Maxwell's equations are reduced to the inhomogeneous Helmholtz equation[29]

$$\left(\nabla^2 - \nabla\nabla^T + k^2\right)\vec{E}(\vec{r}) = -\ddot{F}(\vec{r})\vec{E}(\vec{r}), \qquad (1)$$

where $k$ is the magnitude of the wave vector, $\ddot{F} = k^2\left[\ddot{\varepsilon}(\vec{r})/n_m^2 - 1\right]$ is the scattering potential tensor with $n_m$ as the mounting medium refractive index, and $\vec{E}(\vec{r})$ is the scattered light field. The Helmholtz equation can be equivalently expressed as the Lippmann-Schwinger equation[29,30]

$$\vec{E}(\vec{r}) = \vec{E}_{in}(\vec{r}) + \int \ddot{G}(\vec{r}-\vec{r}')\ddot{F}(\vec{r}')\vec{E}(\vec{r}')d\vec{r}', \qquad (2)$$

where $\vec{E}_{in}$ is the field in the absence of a sample, $\ddot{G} = \left(I + \nabla\nabla^T/k^2\right)G(\vec{r})$ is dyadic Green's function, and $G(\vec{r}) = e^{ik|\vec{r}|}/4\pi|\vec{r}|$ is scalar Green's function. Further, the Lippmann-Schwinger equation can be intuitively understood as Huygens' principle, where the illumination scattered at every sample point generates a spherical wave with an amplitude proportional to the sample's scattering potential multiplied by the field[30].

To perform numerical simulations, this equation is discretized on a voxel grid

$$\vec{E} = \vec{E}_{in} + \vec{G} diag(\vec{F})\vec{E}, \qquad (3)$$

where $\vec{E}$, $\vec{E}_{in}$, and $\vec{F}$ are the discretized versions of $\vec{E}(r)$, $\vec{E}_{in}(r)$, and $\vec{F}(r)$, respectively, and $\vec{G} = U^{\dagger} diag\left[\left(|\vec{q}|^2 - k^2\right)^{-1}\left(1 - \vec{q}\vec{q}^{\dagger}/k^2\right)\right]U$ is the discretized convolution with Green's function, where U is the discrete Fourier transform with frequency coordinates $\mathbf{q}$[25,26]. Finally, the scattered field can be expressed as

$$\vec{E} = \left[1 - \vec{G} diag(\vec{F})\right]^{-1} \vec{E}_{in}. \qquad (4)$$

The matrix inversion on the right-hand side of the equation is a challenge to overcome when computing the scattered field. For a large system, direct inversion of the matrix is not feasible and must be evaluated using iterative algorithms. Here, we used the convergent Born series[31-33] to perform the inversion; however, other approaches are also possible[26,34].

**Dielectric tensor tomography with weak scattering approximation**

Previously, our group demonstrated the use of dielectric tensor tomography (DTT). a polarization-sensitive holographic microscopy technique for the reconstruction of a 3D dielectric tensor [Fig. 2a; see *Methods* for details]. Based on the polarization-resolved holograms of a weakly birefringent sample, DTT reconstructed its 3D dielectric tensor distribution using linear reconstruction theory[23]. This theory assumed the Rytov approximation, wherein the optical path difference of the sample slowly varies in the transverse direction. The derived equation linearly related the unwrapped phase of the transmitted field, $\vec{\psi}$, with the 3D dielectric tensor in Fourier space as

$$\vec{\tilde{F}}\left(k_x - k_{i,x}, k_y - k_{i,y}, k_z - k_{i,z}\right)\vec{p}_i \oslash \left(-2ik_z \vec{p}_i\right) = \vec{\tilde{\psi}}\left(k_x - k_{i,x}, k_y - k_{i,y}\right). \qquad (5)$$

where ~ on the top hat denotes the Fourier transform signal, $\mathbf{q} = (k_x, k_y, k_z)$ is the coordinate in the Fourier space, and $\mathbf{k}_{in} = (k_{i,x}, k_{i,y}, k_{i,z})$ and is the wavevector and 3D polarization of the *i*-th incident plane wave, respectively. Further, $|\mathbf{k}_{in}| = k$ and $\oslash$ is Hadamard division. To determine every element of the 3 × 3 symmetric dielectric tensors from two polarization-resolved holograms, the previously proposed method required additional measurement with a slightly tilted illumination angle. However, implementing this procedure is challenging and prone to noise owing to the requirement of precise control of the illumination beam and differential measurements. Moreover, the algorithm is inaccurate in the case where sample-induced birefringence and scattering are significant.

**Reconstruction algorithm of highly scattering dielectric tensor**

Based on the complete understanding of the vectoral diffraction process, the main goal of our study can be addressed, that is, imaging highly scattering birefringent samples. This is referred to as the inverse problem, and it aims to retrieve the dielectric tensor from a known illumination ($\vec{E}_{in}$) and scattered field ($\vec{E}$). We developed a multiple scattering algorithm to find the dielectric tensor ($\ddot{\varepsilon}$) wherein the simulated scattered fields most closely matched the imaged fields [Fig. 2b]. This goal can be mathematically expressed as the following minimization problem:

$$\ddot{\varepsilon} = \min_{\ddot{\varepsilon}'} c(\vec{\varepsilon}') = \min_{\ddot{\varepsilon}'} \frac{1}{2} \left\| \vec{y}'(\vec{\varepsilon}') - \vec{y} \right\|^2. \tag{6}$$

This is the square norm of the difference between the measured and simulated scattered fields $\vec{y}$ and $\vec{y}'(\vec{\varepsilon}')$, respectively. To perform the minimization, the gradient of the minimized function must be computed to enable the use of a gradient descent algorithm.

To compute the gradient, we used the scattering potential ($\vec{F}$) instead of the dielectric tensor ($\ddot{\varepsilon}$) to simplify the derivations. When the measured field is obtained experimentally, the simulated scattered field can be expressed using equation Eq. (1).

$$\vec{y}'(\vec{F}') = \vec{P} \left[ 1 - \vec{G} diag(\vec{F}') \right]^{-1} \vec{E}_{in}, \tag{7}$$

where $P = U^{\dagger} \text{diag}\left[ \text{rect}(\lambda q_{xy}/2\pi \text{NA}) \exp(-i q_z z) \right] U \text{diag}\left[ \delta(z - z_{far}) \right]$ is the far-field projection operator[25] expressing spatial filtering owing to the limited numerical aperture of the system. Further, the cost function gradient is equivalent to

$$\nabla_{\vec{F}'}\left[c\left(\vec{F}'\right)\right] = \left[\left(1-\vec{G}diag\left(\vec{F}'\right)\right)^{-1} diag\left(\overline{E}_{in}\right)\right]^{*} \odot_{r} \left[\left(1-\vec{G}diag\left(\vec{F}'^{\dagger}\right)\right)^{-1} \vec{G}P^{\dagger}\Delta\vec{y}'\left(\vec{F}'\right)\right]_{ij}^{T}, \quad (8)$$

where $A_{ij}^T$ denotes the transpose of A along the 3 × 3 tensor dimension only, $\odot_r$ denotes the Hadamard product along the spatial dimension with a matrix product along the 3 × 3 tensor dimension, and * denotes the element-wise complex conjugate (detailed derivation has been provided in *Methods*).

A comparison of Eqs. (1) and (3) indicate that the gradient can be computed using two sequences of vectoral forward computations. The left term corresponds to the scattering simulation by the dielectric tensor estimate of the sample, and the right term corresponds to the scattering simulation of the phase conjugate of the difference between the measured and scattered fields. This last step can be understood as an error backpropagation or as the time reversal of the error owing to the use of phase conjugation on the error field[35,36].

The pseudocode of the reconstruction algorithm is presented in Table 1. During the gradient descent step, we improved the quality of the reconstructed tomograms by imposing constraints related to prior knowledge of the samples. In this study, we used a total variation regularization algorithm combined with a fast iterative shrinkage-thresholding algorithm (TV-FISTA)[37,38]. In our previous study, we imposed the algorithm on individual tensorial elements and circumvented the missing cone problem[39]. However, the method of independent regularization caused errors in the reconstructed directors in the presence of experimental noise. To mitigate this problem, we improved the regularization algorithm by imposing a similarity between the principal RIs by constraining the total birefringence (TB), which is defined as

$$TB\left(\vec{A}\right) = 3 \cdot tr\left(\vec{A}^{\dagger}\vec{A}\right) - tr\left(\vec{A}^{\dagger}\right)tr\left(\vec{A}\right) = \left|n_1^2 - n_2^2\right|^2 + \left|n_2^2 - n_3^2\right|^2 + \left|n_1^2 - n_3^2\right|^2, \quad (9)$$

where $n_i^2$ are the eigenvalues of $\vec{A}$, with $n_o$ corresponding to the principal RIs of the dielectric tensor. The dielectric tensor is also known to be symmetrical; therefore, this constraint was imposed during the optimization process. Details of the implementation of the constraints are presented in the *Methods* section.

**Experimental validation using synthetic liquid crystals**

We experimentally validated the proposed reconstruction method by imaging a synthetic liquid crystal with a known structure. We chose to use a liquid-crystalline microsphere made of sodium dodecyl sulfate (LCSDS) in a 1.52 ultraviolet-cured medium. The LCSDS microsphere exhibits a radial arrangement of directors and was used as a reference sample. The amount of birefringence was defined as the $\frac{\sqrt{(n_1^2 - n_2^2)^2 + (n_1^2 - n_3^2)^2 + (n_3^2 - n_2^2)^2}}{\sqrt{8} \cdot n_m}$.

We compared the reconstruction results of three different methods: (i) Rytov approximation, (ii) TV-FISTA with single-scattering approximation, and (iii) our proposed method [Fig. 3a]. All reconstruction methods showed consistent imaging results, with similar RI directions and birefringence. However, we observed that the proposed method outperformed the other methods, particularly by providing the most homogeneous birefringence limited to the spatial bandwidth and expected spherical shape in the vertical direction. The other reconstruction methods suffered from elongation in the axial direction. These results were unexpected because we reduced the RI contrast of the sample to match the weakly scattering condition for DTT to the best extent possible. This suggests that inaccurate scattering models suffer from both the missing cone problem and multiple scattering artifacts even after using a regularization algorithm. Thus, our inverse model can significantly improve the reconstruction quality in DTT even in the case of small RI contrast.

**3D reconstruction of the highly scattering and birefringent biostructure**

A promising application of DTT is in the analysis of molecular alignment in biological specimens. However, most birefringent biological specimens are too optically heterogeneous for imaging. A highly scattered starch sample was prepared in a mounting medium with a refractive index of 1.43.

The results indicated a significant improvement in reconstruction quality when using the proposed method [Fig. 3b]. The single-scattering model exhibited noisy artifacts in the 3D retrieved birefringence and director maps. In contrast, the proposed method clarified that the directors aligned perpendicularly to the surface of the starches while measuring homogeneous birefringence. Remarkably, the perpendicular alignment of the directors was also visible in the axial cross-section, even when the starch was non-spherical. The improved result is consistent with the "growth ring" structure observed by X-ray microscopy, where the starch grows by alternating amorphous and crystalline layers, wherein the principal axis is perpendicular to the surface[40]. Additionally, the hilum, the singular center of the largest starch, was more clearly visualized using the multiple scattering model.

## Discussion

We developed a reconstruction method to image the 3D dielectric tensors of highly scattered samples. The original DTT reconstruction method although computationally fast, assumed weakly scattering approximations and required angularly differential measurement, which limited its applications. Although our method is slower than the previous algorithm, it improved reconstruction accuracy by considering multiple light scattering. This was experimentally validated by reconstructing synthetic liquid crystal droplets and biological starch samples.

The developed theory was numerically implemented using the convergent Born series, which is among the most efficient forward solvers for electromagnetic waves. However, our method is compatible with other electromagnetic solvers and can be tested in immediate follow-up studies. In addition, the proposed technique can be further improved by reducing the computational burden and developing methods to enable convergence to the global minimum. Furthermore, a synergistic approach between optical theory and computational modeling may extend our method to the high-throughput analysis of biological tissues and multicellular interactions.

## Materials and Methods

**Optical setup.** The setup was based on a polarization-sensitive Michelson interferometer [Fig. 2a]. A 532 nm continuous wave laser (Cobolt AB) was split into a sample and a reference arm using a beam splitter. In the former, a digital micromirror device (DLi 4130, Digital Light Innovations) and liquid-crystal retarder (LCC1223-A, Thorlabs) were used to control both the illumination angle and polarization states of the incident light. A water-immersion condenser (UPLSAPO60XW, NA = 1.2, Olympus) and objective lens (UPLSAPO60XO, NA = 1.42, Olympus) was used to illuminate the sample and collect the diffracted light field. Further, the diffracted light field was split using a polarizing beam splitter and interfered with the reference beam on two cameras (Lt425R, Lumenera) to form two polarization-sensitive off-axis holograms.

**Derivation of error gradient.** A detailed derivation of Eqs. (6) to (8) are based on matrix calculus theory. Using Eqs. 247, 136, and 232 from the matrix cookbook[41] and the fact that $\vec{y}'(\vec{F}')$ is analytic in $\vec{F}'$ because $\vec{G}diag(\vec{F}')$ is generally not equivalent to an identity matrix[30], we computed the derivative of the error function concerning the tensorial element, $F_{ij}$, using the chain rule as follows:

$$\frac{\partial}{\partial F_{\vec{r}ij}} c(\vec{F}') = \frac{\partial}{\partial F_{\vec{r}ij}} \left(\Delta y'(\vec{F}')\right)^\dagger \Delta y'(\vec{F}')$$
$$= \left[ P\left(1-\vec{G}diag(\vec{F}')\right)^{-1} \vec{G}diag(\delta_{\vec{r}ij})\left(1-\vec{G}diag(\vec{F}')\right)^{-1} diag(\vec{E}_{in}) \right]^\dagger \Delta y'(\vec{F}'), \quad (10)$$

where $ij$ denotes the 3 × 3 tensor dimensions and $\vec{r}$ denotes the spatial dimension. Using the identity $\left[1-\vec{G}diag(\vec{F}')\right]^{-1}\vec{G} = \vec{G}\left[1-diag(\vec{F}')\vec{G}\right]^{-1}$ and the Hermitian property of Green's function ($\vec{G}^\dagger = \vec{G}$), the equation is converted to

$$\frac{\partial}{\partial F_{\vec{r}ij}} c(F') = \left[\left(1-\vec{G}diag(\vec{F}')\right)^{-1} diag(\vec{E}_{in})\right]^\dagger diag(\delta_{\vec{r}ij}) \left[1-\vec{G}diag(\vec{F}'^\dagger)\right]^{-1} \vec{G}P^\dagger \Delta \vec{y}'(\vec{F}'). \quad (11)$$

Using the properties of the Kronecker-delta tensor, we can reorganize the above equation to Eq. (8).

**TV-FISTA for the single-scattering model.** To reduce the artifacts caused by experimental noise and the missing-cone problem, we implemented a TV-FISTA regularization algorithm. In the single-scattering model, the algorithm minimizes the following error function concerning each tensorial element.

$$\min_{\vec{F}} \left\| A(\vec{F}) - y \right\|^2 - \tau \, \text{TV}(\vec{F}), \quad (12)$$

where $y$ is the measured field phase, $A(\vec{F})$ corresponds to Eq. (5), $\tau$ is the TV parameter, and $TV(x) = \|\nabla x\|_2$ is the regularization term [39]. Regularization was applied independently for each tensorial element. The step sizes, TV parameters, and several outer and inner iterations are described in the following section.

**Total birefringence regularization for the multiple-scattering model.** Equation (2) defines the total birefringence (TB) of a 3 × 3 tensor, $A = \begin{pmatrix} A_1 & A_2 & A_3 \\ A_4 & A_5 & A_6 \\ A_7 & A_8 & A_9 \end{pmatrix}$. Its global minimum value is zero when a sample is optically isotropic and $n_1^2 = n_2^2 = n_2^3$. Moreover, TB is a convex function[42], which can be proven by investigating its Hessian matrix

$$\frac{\partial^2 TB}{\partial A_i \partial A_j} = \begin{pmatrix} 4 & 0 & 0 & 0 & -2 & 0 & 0 & 0 & -2 \\ 0 & 6 & 0 & 0 & 0 & 0 & 0 & 0 & 0 \\ 0 & 0 & 6 & 0 & 0 & 0 & 0 & 0 & 0 \\ 0 & 0 & 0 & 6 & 0 & 0 & 0 & 0 & 0 \\ -2 & 0 & 0 & 0 & 4 & 0 & 0 & 0 & -2 \\ 0 & 0 & 0 & 0 & 0 & 6 & 0 & 0 & 0 \\ 0 & 0 & 0 & 0 & 0 & 0 & 6 & 0 & 0 \\ 0 & 0 & 0 & 0 & 0 & 0 & 0 & 6 & 0 \\ -2 & 0 & 0 & 0 & -2 & 0 & 0 & 0 & 4 \end{pmatrix}, \quad (13)$$

which has nonnegative eigenvalues (0 and 8 degenerate for 6s). The minimization problem can be solved using a gradient descent algorithm. We implemented it using the TV-MFISTA algorithm[38]. We further note that the Lipschitz constant and gradient required for MFISTA optimization are

$$\nabla TB(A) = 6A - 2tr(A)$$
$$L(TB) = 6 \quad (14)$$

In all our reconstructions, the total birefringence was applied with a factor of 0.001, while the total variation was applied with a factor of 0.01. For linear reconstruction, the total variation was applied with a slightly larger factor of 0.04 owing to the use of Fourier space reweighting. Further, for the MFISTA algorithm, we used 800 inner iterations and 100–80 outer iterations for the bead and starch samples, respectively.

Using the symmetric property of the dielectric tensor, we can impose an additional constraint on the dielectric tensor using the proximal map

$$prox_t(g)(x) = \arg\min_u \left[ g(u) + \frac{1}{2t}\|x - u\|^2 \right], \quad (15)$$

where $g(u)$ is the indicator function of symmetric $u$. By setting $u = \begin{pmatrix} u_{11} & u_{12} & u_{13} \\ u_{12} & u_{22} & u_{23} \\ u_{13} & u_{23} & u_{33} \end{pmatrix}$ and

$x = \begin{pmatrix} x_{11} & x_{12} & x_{13} \\ x_{21} & x_{22} & x_{23} \\ x_{31} & x_{32} & x_{33} \end{pmatrix}$, the minimum condition satisfies $\frac{\partial}{\partial u_{ij}}\left[ g(u) + \frac{1}{2t}\|x - u\|^2 \right] = 0$. The result obtained is

$$prox_t(g)(x) = \begin{pmatrix} x_{11} & \frac{x_{12} + x_{21}}{2} & \frac{x_{13} + x_{31}}{2} \\ \frac{x_{12} + x_{21}}{2} & x_{22} & \frac{x_{23} + x_{32}}{2} \\ \frac{x_{13} + x_{31}}{2} & \frac{x_{23} + x_{32}}{2} & x_{33} \end{pmatrix}. \quad (16)$$

Note that the imaginary part of the tensor was set as a scalar.

**Acknowledgements**

This work was supported by KAIST UP program, BK21+ program, Tomocube Inc., National Research Foundation of Korea (2015R1A3A2066550, 2022M3H4A1A02074314), KAIST Institute of Technology Value Creation, Industry Liaison Center (G-CORE Project) grant funded by the Ministry of Science and ICT (N11210014, N11220131) and Institute of Information & communications Technology Planning & Evaluation (IITP; 2021-0-00745) grant funded by the Korea government (MSIT)

**Conflict of interest**

The authors declare that there are no conflicts of interest related to this article.

**Author contributions**

H.H. performed and analysed the experiments, and developed a theory. M.L. developed a theory. S.S. provided analytical tools and supported the experiments. Y.P. supervised the project. All authors wrote the manuscript.

**Figures with legends**

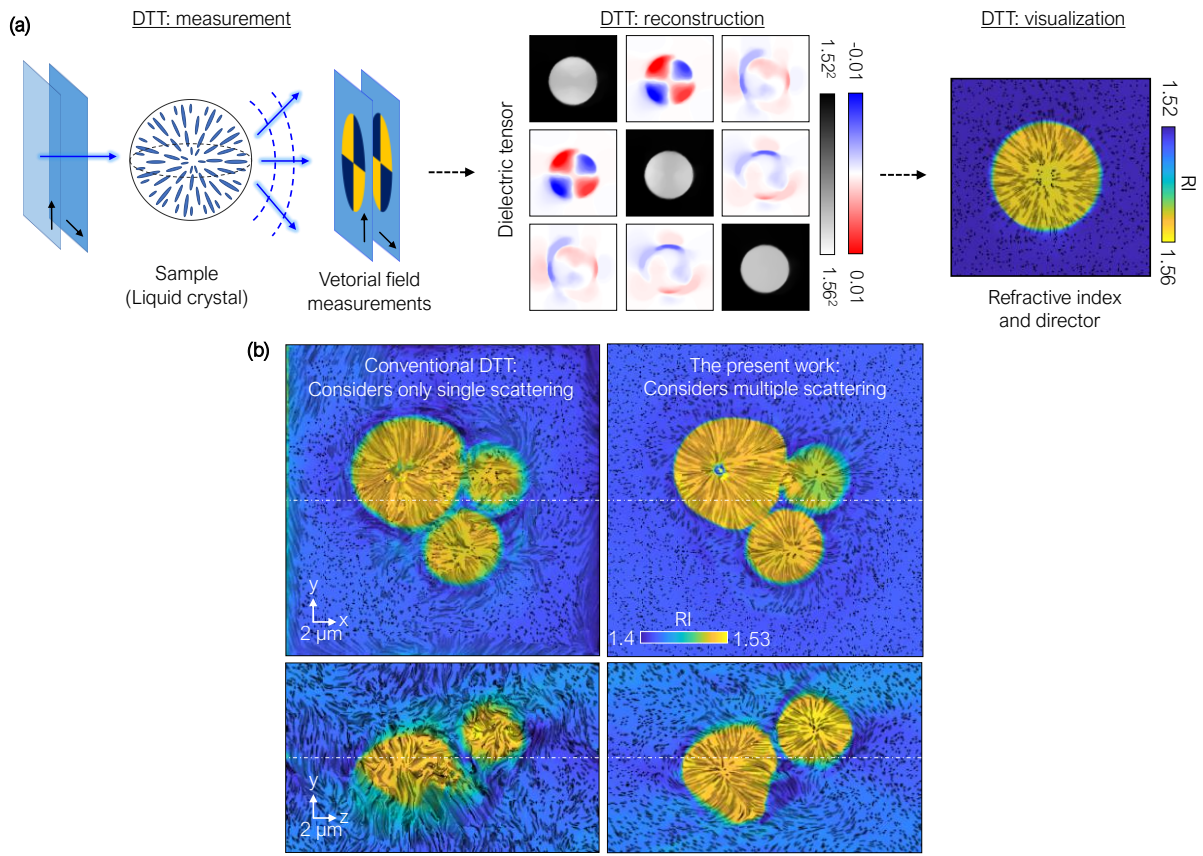

**Figure 1 | Schematic of the proposed reconstruction method.** (a) Dielectric tensor tomography of a radially distributed liquid crystal droplet, which is reconstructed from the measured multiple vectorial field images. The dielectric tensor can be decomposed using singular value decomposition into 3 orthogonal polarization directions and 3 refractive indices for each of the corresponding polarizations. Then, the dielectric tensor can be converted into the refractive index, director, and birefringence for visualization. (b) Representative sample (potato starch) reconstructed using both the single and multiple scattering models. Birefringence direction is consistent with the growth ring structure of starches.

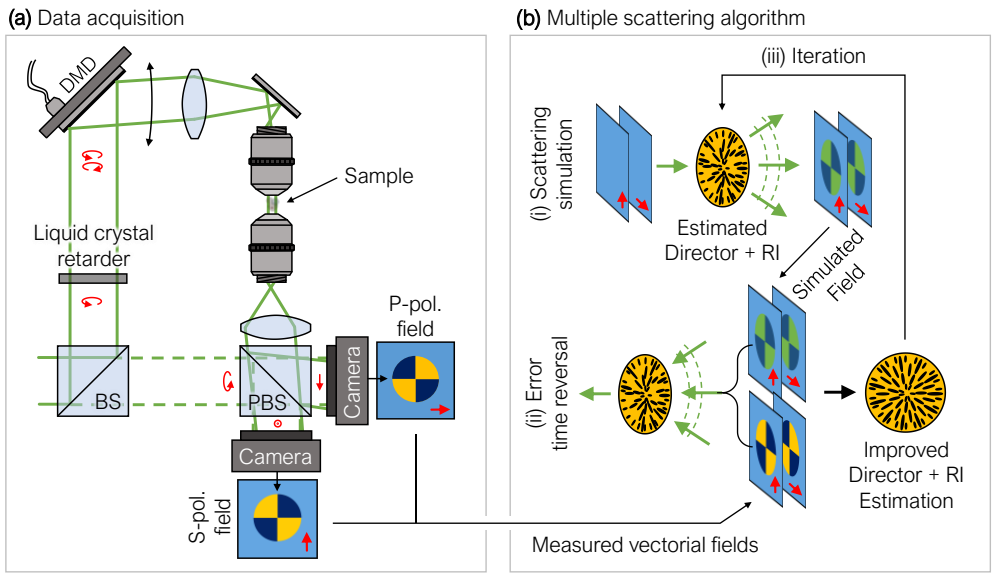

**Figure 2 | Reconstruction algorithm principle.** (a) Optical setup. DMD: digital micromirror device, BS: beam splitter, PBS: polarized beam splitter. (b) Dielectric tensor reconstruction algorithm (i) simulation of the scattering through the sample (ii) backpropagation of the error (iii) update of the refractive index.

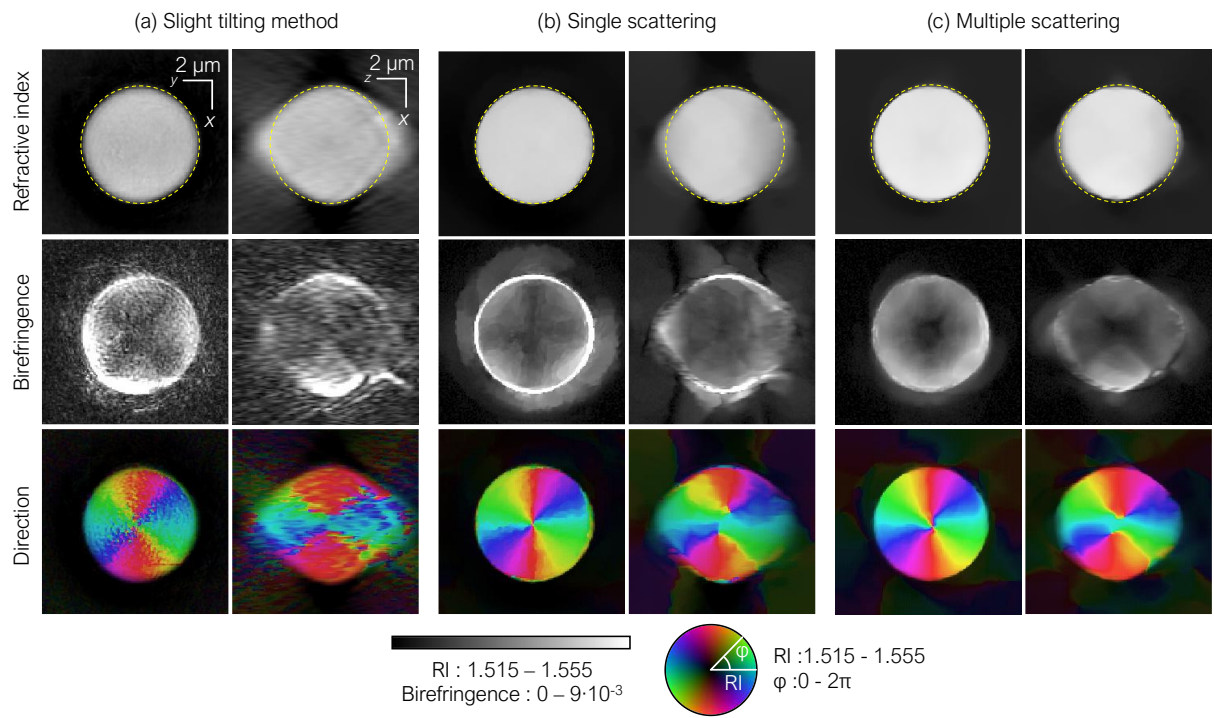

**Figure 3 | Reconstruction results of an LCSDS microsphere** using (a) slight tilting, (b) single scattering DTT, and (c) multiple scattering DTT method.

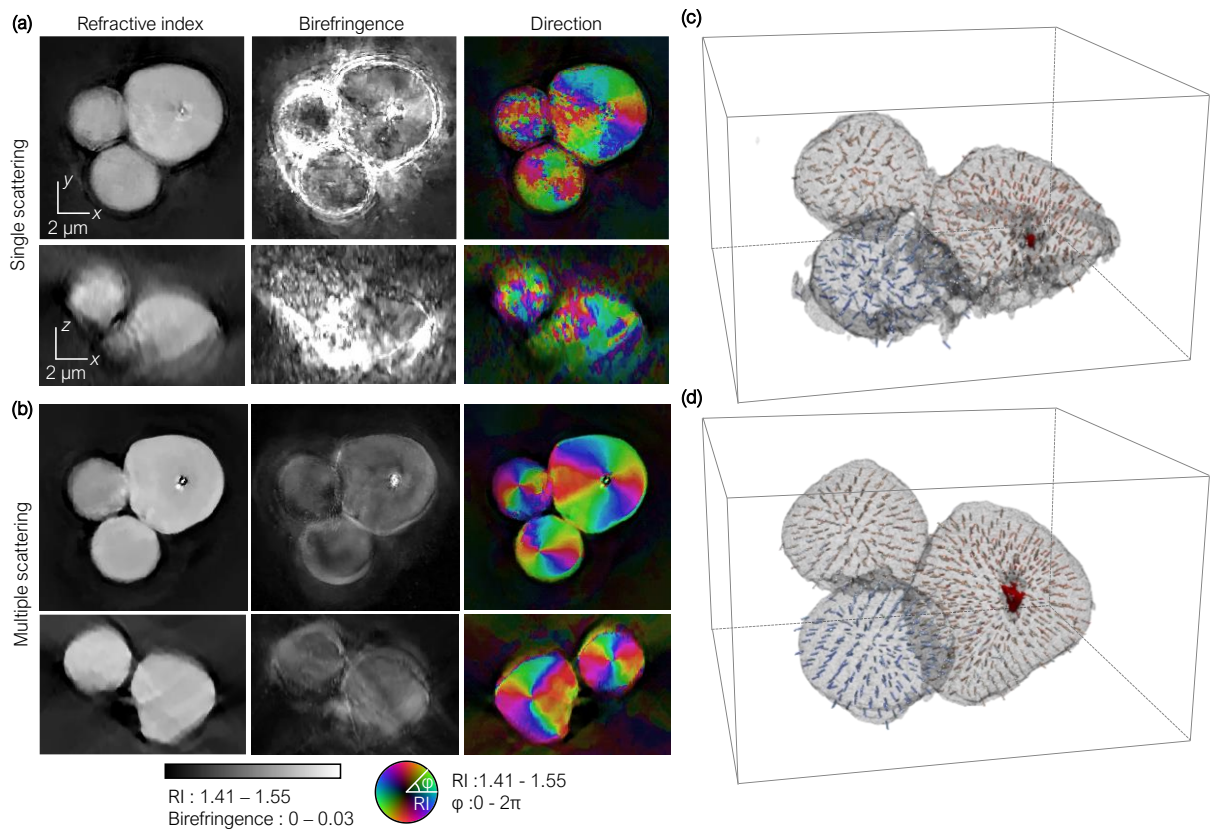

**Figure 4 | Reconstruction results of multiple starches.** (a-b) Refractive index, birefringence, and directors of the sample using (a) the single scattering and (b) multiples scattering DTT method. (c-d) Corresponding three-dimensional renderings of principle refractive index values and directors.

|  | Optical diffraction tomography 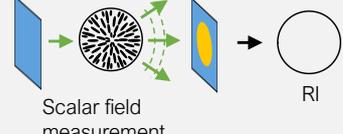 Scalar field measurement → RI | Dielectric tensor tomography 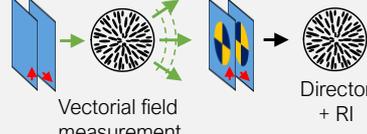 Vectorial field measurement → Director + RI |
|---|---|---|
| Single scattering algorithm 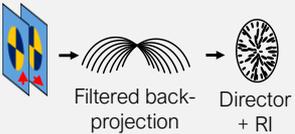 Filtered back-projection → Director + RI | Paper : "Three-dimensional structure determination of semi-transparent objects from holographic data" E. Wolf<br><br>Advantages :<br>- Very fast<br><br>Disadvantages :<br>- Does not measure birefringence<br>- Does not corrects multiple scattering | Paper :"Tomographic measurement of dielectric tensors at optical frequency" S.Shin et al.<br><br>Advantages :<br>- Fast<br>- Measure birefringence<br><br>Disadvantages :<br>- Does not corrects multiple scattering |
| Multiple scattering algorithm 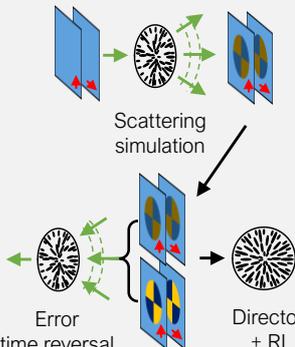 Scattering simulation, Error time reversal → Director + RI | Paper : "Learning approach to optical tomography" U.S. Kamilov et al.<br><br>Advantages :<br>- Corrects multiple scattering<br><br>Disadvantages :<br>- Slow<br>- Does not measure birefringence | Paper : The present work<br><br>Advantages :<br>- Measures birefringence<br>- Corrects multiple scattering<br><br>Disadvantages :<br>- Slow |

**Figure 5 | Comparison between different imaging modalities and algorithms.**

| Reconstruction algorithm |
|---|
| Y = measured fields, ε' = dielectric tensor estimate, $E_{in}$ = illumination field |
| **While** not converged |
|     [y', $E_{3D}$] = simulate($E_{in}$, ε') |
|     [ ~, $E_{err}$] = simulate(y'-y, flip(ε'$^{*T}$, 3)) |
|     $E_{err}$(:,:,:,3,:) = -$E_{err}$(:,:,:,3,:) |
|     gradient = i $E_{3D}^*$ $\odot_x$ flip($E_{err}$, 3) $^T{}_{ij}$ |
|     ε' = optimize(ε', gradient) |

**Table 1 | Pseudocode.**

[$E_{far}$, $E_{3D}$] = simulate($E_{in}$, F) is the scattering simulation giving the scattered field $E_{far}$ at the camera plane and the 3D field inside the object $E_{3D}$ for an object of scattering potential F and illumination $E_{in}$. F = optimize(F, gradient) is one step of gradient-based optimization; in our case, the FISTA algorithm[37,38].